\documentclass[aps,physrev,twocolumn,groupedaddress]{revtex4-2}
\usepackage{amsmath}
\usepackage{graphicx}
\usepackage{comment}

\begin{document}


\title{Experimental Reconstruction of Source 4D Phase Space Without Prior Knowledge of Transfer Matrix}


\author{Charles Zhang $^1$, Elena Echeverria$^1$, Abigail Flint$^1$, William H, Li $^2$, Christopher M. Pierce $^3$, Alice Galdi $^4$, Chad Pennington $^5$, Adam Bartnik$^1$, Ivan Bazarov$^1$, Jared Maxson$^1$}

\affiliation{$^1$Cornell Laboratory for Accelerator-Based Sciences and Education, Cornell University, Ithaca, NY, USA}
\affiliation{$^2$Brookhaven National Laboratory, Upton, NY, USA}
\affiliation{$^3$Enrico Fermi Institute, University of Chicago, Chicago, IL 60637, USA}
\affiliation{$^4$Department of Industrial Engineering, University of Salerno, Fisciano (SA) Italy}
\affiliation{$^5$University of California, Los Angeles, CA, USA}


\date{\today}

\begin{abstract}

We experimentally demonstrate a method for reconstructing the transverse 4D phase space of an electron beam at the time of emission from downstream diagnostics of the 4D phase space. This method does not rely on detailed knowledge of the beamline transport, besides assuming that linearity and symplecticity are satisfied. We apply this method to measure the transverse position and momentum phase space of electrons emitted from a spatially-structured alkali-antimonide cathode. This method can uncover local correlations between emission location and momentum spread. We formulate this method analytically and investigate resolution limits.

\end{abstract}


\maketitle

\section{Introduction}
Understanding the physics of photoemission is critical to many applications of particle beams. Cathode photoemission performance, often characterized by the mean transverse energy (MTE) and quantum efficiency, determines (in part) the maximum achievable brightness of a beam in a linear accelerator for which Liouville's theorem applies to the 6D phase space distribution. Brightness is a key metric for various applications such as ultrafast electron microscopy \cite{MeV_UED,li2022kiloelectron}, x-ray free electron lasers \cite{emma2010first, bostedt2016linac}, and colliders \cite{meshkov2019luminosity,2015optimizingluminosity}. Brightness can be defined in several ways; nearly all definitions of brightness include a factor that depends on the transverse 4-D phase space distribution. Our work focuses on a new method to measure the source 4D transverse phase space distribution.

There has been much work done, both theoretically \cite{DS-theory, vecchione2013quantum,sid_surfacedisorder, gevork_surfacedisorder,saha2023theory} and experimentally \cite{alice_singlecrystalcathode,karkare2020ultracold,pennington2025structural,parzyck2023atomically}, to study the QE and MTE of cathodes of various materials and surface preparations in accelerator contexts, in high surface field and charge/current regimes. However, experimental work has yet to bridge the gap with the theory in exploring the photoemission process at a microscopic level beyond ensemble quantities such as QE and MTE. Specifically, the full distribution of source momentum, as well as local correlations with surface deformities (physical and chemical), is not frequently measured in accelerators, which can hamper the prediction of beam dynamics and comparison to photoemission theories. As such, there is a need for an accessible experimental diagnostic that can measure the beam in an accelerator and reconstruct its source 4D phase space. 

The MTE of a cathode is typically measured through several techniques \cite{lee2015review}, such as scans of electromagnetic optics (solenoid, quadrupole, accelerating field/voltage) \cite{bazarov2008thermal,wiedemann2015particle,lee2015review}, or aperture/pepperpot measurements \cite{Matt_4demittance}. These measurements are either performed in actual accelerators or in smaller characterization chambers. Often, these methods do not directly measure the source momentum distribution, but instead measure the normalized beam emittance $\epsilon$. The MTE is then given by $\text{MTE} = mc^2(\epsilon/\sigma_x)^2$, where $\sigma_x$ is the rms transverse size of the beam in real space at the cathode.  To do this, scans of optics (like solenoid scans) typically require accurate fieldmaps that describe the beamline elements throughout the beam transport. Uncertainties in the transport directly result in uncertainties in emittance/MTE. In addition, it can be non-trivial to measure the rms transverse size of the beam, $\sigma_x$, at the cathode, e.g., in cases where the quantum efficiency or drive laser intensity varies rapidly in space, or in cases with significant multiphoton photoemission \cite{musumeci2010multiphoton,knill2023effects}.

Some advanced probes, such as photoemission electron microscopy (PEEM) \cite{kachwala2023demonstration,feng2007photoemission} or angle-resolved photoemission spectroscopy \cite{hong2023anomalous}, do measure the full momentum distribution of the photoemitted electrons. However, most PEEMs require a complex set of lens calibration procedures, which can be expensive or time-consuming to set up. The hemispherical analyzers used in ARPES also typically require eV-scale electron kinetic energies, which, for a photocathode driven by visible or near/mid-ultraviolet light, requires post-acceleration, which invariably adds lensing effects that must be accounted for \cite{hong2023anomalous}. Despite these challenges, PEEM and ARPES provide the highest resolution images of photoemission momentum. Yet, these excellent probes are not likely to be available in the accelerator itself, and do not trivially (i.e., without tomographic techniques) resolve local-position-momentum correlations at the source.

In this paper, we demonstrate a method where downstream 4D phase space measurements, achieved via aperture scans, can be used to reconstruct the source transverse phase space without prior knowledge of the transfer matrix. We deduce the transfer matrix by measuring the experimentally accessible matrix elements (for us, these are the position-position and position-momentum elements), and then exploiting the symplectic condition to infer the values of the unknown transfer matrix elements (in our case the momentum-position and momentum-momentum elements). Here, the momentum-position and momentum-momentum transfer matrix elements are experimentally inaccessible since we do not have a means to deterministically vary the momentum of the beam at emission. Once the transfer matrix is known, the source phase space can be determined via application of the inverse transfer matrix to the measured downstream phase space.  

We first discuss the theory behind this source reconstruction method in Sec. \ref{sec:theory}, our experimental setup in Sec. \ref{sec:exp_setup}, and then discuss the practical implementation of this method and error considerations in Sec. \ref{sec:results} and  \ref{sec:error}.

\section{Theory} \label{sec:theory}
\begin{figure*}
    \centering
    \includegraphics[width=0.8\linewidth]{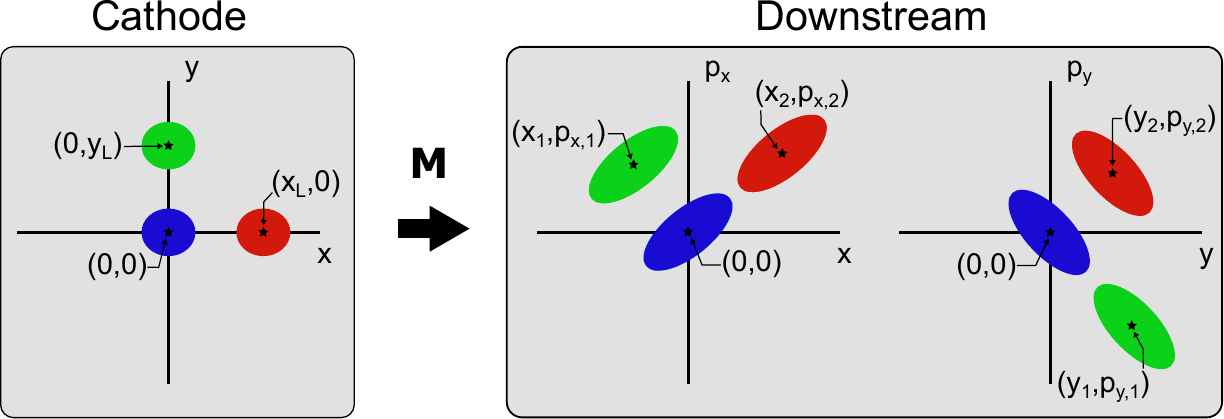}
    \caption{A cartoon demonstrating the measurement of the partial transfer matrix. At the cathode (left), the beam is moved on the emission surface to positions $(x_L, 0)$ and $(0,y_L)$. At a downstream diagnostic (right), after evolving with the transfer matrix $\mathbf{M}$, the 4D phase space of the beam is measured and labeled by its centroids (indicated by stars). Each phase space is measured with the beam emitted at the correspondingly colored cathode position.}
    \label{fig:theory}
\end{figure*}

A linear transfer matrix, $\mathbf{M}$, is often a good approximation for relating the source position and momentum of a particle to its final position and momentum downstream near the design trajectory of a beamline,
\begin{align}
\begin{split}
    \begin{pmatrix} x \\ p_x \\ y \\ p_y \end{pmatrix}_f  = \mathbf{M} \begin{pmatrix} x \\ p_x \\ y \\ p_y \end{pmatrix} _i 
    \\
    \mathbf{M}  = \begin{pmatrix} m_{11} & m_{12} & m_{13} & m_{14} \\ m_{21} & m_{22} & m_{23} & m_{24} \\ m_{31} & m_{32} & m_{33} & m_{34} \\ m_{41} & m_{42} & m_{43} & m_{44} \\  \end{pmatrix}
    \end{split}
    \label{eq:transfer_matrix}
\end{align}
where $x, y$ are transverse coordinates and $p_x, p_y$ are transverse (physical) momenta. In this work, we restrict consideration to 4 dimensions, but many of our results can be generalized to 6D provided a means to measure some of the longitudinal transfer matrix elements.

According to Hamiltonian mechanics, the transfer matrix must satisfy the symplectic condition,
\begin{align}
\begin{split}
    \mathbf{M^{-1}} & =  -\Omega \mathbf{M^T} \Omega 
    \\
   &  = 
    \begin{pmatrix}
m_{22} && -m_{12} && m_{42} && -m_{32}\\
-m_{21} && m_{11} && -m_{41} && m_{31}\\
m_{24} && -m_{14} && m_{44} && -m_{34}\\
-m_{23} && m_{13} && -m_{43} && m_{33}\\
\end{pmatrix}
\\
\Omega & = \begin{pmatrix}
        0 & 1 & 0 & 0 \\ -1  & 0 & 0 & 0 \\ 0 & 0 & 0 & 1 \\ 0 & 0 & -1 & 0
    \end{pmatrix}
\end{split}
    \label{eq:symplectic_condition}
\end{align}
Note that the symplectic condition constrains the inverse significantly: it is a simple rearrangement and sign modification of the original transfer matrix in Eq. \ref{eq:transfer_matrix}. This, as we will show, allows us to calculate the full source momentum distribution without full knowledge of the transfer matrix. 

In principle, we can measure components of the transfer matrix by linearly fitting the final position and momentum of a particle measured at a beamline diagnostic to its source position and momentum. In practice, for a beam of finite size, its centroid in 4D phase space transforms like a single particle. To determine transfer matrix elements from a particle source,  we often only have control over the source position centroid. In the case of a photocathode, this can be done by moving the laser driving photoemission a known displacement on the cathode surface.

For each spatial displacement of the emission site, the \emph{phase space displacement} of the centroid (a 4D vector) of the beam at the diagnostic can be measured by a phase space diagnostic. For example, as shown in Fig. \ref{fig:theory}, we measure phase space centroids of $(x_1, p_{x,1}, y_1,p_{y,1})$ and $(x_2, p_{x,2}, y_2,p_{y,2})$ for a laser displacement of  $(y_L,0)$ and $(x_L,0)$,  respectively.   

Altogether, this allows us to determine the first and third columns of the transfer matrix, which will be referred to as the partial transfer matrix,
\begin{align*}
    \begin{pmatrix} m_{11}& m_{13}\\ m_{21} & m_{23} \\ m_{31} & m_{33}  \\ m_{41}  & m_{43} \\  \end{pmatrix} = \begin{pmatrix} x_2/x_L & x_1/y_L \\ p_{x,2}/x_L  & p_{x,1}/y_L \\ y_2/x_L & y_1/y_L   \\ p_{y,2}/x_L  & p_{y,1}/y_L   \end{pmatrix} 
\end{align*}
Note that this requires us to assume that the source momentum centroid does not vary with emission location. Our formalism does allow the momentum \emph{width} (e.g, local MTE) to vary across the emitter surface. Also note that all matrix elements of the form $m_{nl}$ where $l$ is a momentum index (i.e. $m_{nl}$ acts on the source momentum $p_{x,i}, p_{y,i}$), and $n\in[1,4]$ remain unknown, as we typically do not have a means to change the source momentum centroid systematically.

Even without the full transfer matrix, applying an inverse \emph{partial} transfer matrix given by the symplectic condition in Eq. \ref{eq:symplectic_condition} to a point in the measured 4D phase space, we can then calculate its source momentum $p_{x,i}$ and $p_{y,i}$,

\begin{align}
\label{eq:IMPS}
\begin{split}
 \begin{pmatrix}
        p_{x} \\ p_y
    \end{pmatrix}_i & = \begin{pmatrix}
-m_{21} && m_{11} && -m_{41} && m_{31}\\
-m_{23} && m_{13} && -m_{43} && m_{33}\\
\end{pmatrix} \begin{pmatrix} x \\ p_x \\ y \\ p_y \end{pmatrix}_f  
\\
& = \begin{pmatrix}
\frac{-p_{x,2}}{x_L} && \frac{x_2}{x_L}  && \frac{-p_{y,2}}{x_L} && \frac{y_2}{x_L} \\
\frac{-p_{x,1}}{y_L} && \frac{x_1}{y_L} && \frac{-p_{y,1}}{y_L} && \frac{y_1}{y_L} \\
\end{pmatrix} \begin{pmatrix} x \\ p_x \\ y \\ p_y \end{pmatrix}_f
\end{split}
\end{align}
Given a measured 4D phase space, this directly yields the source momentum phase space of the beam, from which the MTE can be calculated as $\sqrt{\left< p_xp_x \right>_i\left< p_yp_y \right>_i - \left< p_x p_y \right>_i^2 }/m$, where $m$ is the particle mass. Under the typical assumptions for disordered polycrystalline cathodes, where the momentum space is round, $\left< p_xp_x\right>_i$ and  $\left< p_yp_y\right>_i$ are equivalent and $\left< p_x p_y\right>_i=0$. Our expression for MTE can then be reduced to the usual formula of $\text{MTE} = \left< p_xp_x\right>_i/m$. 

So far we have reconstructed the source tranverse momentum distribution. With a few additional assumptions,  we can also solve for the full $4\times4$ transfer matrix, and hence reconstruct the source 4D phase space. We first generate a system of non-linear equations from expanding a slightly different form of the symplectic condition in Eq. \ref{eq:symplectic_condition}, $\mathbf{M}\Omega\mathbf{M}^T = \Omega$,
\begin{widetext}
    \begin{align*}
        \begin{pmatrix} m_{11} & m_{12} & m_{13} & m_{14} \\ m_{21} & m_{22} & m_{23} & m_{24} \\ m_{31} & m_{32} & m_{33} & m_{34} \\ m_{41} & m_{42} & m_{43} & m_{44} \\  \end{pmatrix}  \begin{pmatrix}
        0 & 1 & 0 & 0 \\ -1  & 0 & 0 & 0 \\ 0 & 0 & 0 & 1 \\ 0 & 0 & -1 & 0
    \end{pmatrix}  \begin{pmatrix} m_{11} & m_{21} & m_{31} & m_{41} \\ m_{12} & m_{22} & m_{32} & m_{42} \\ m_{13} & m_{23} & m_{33} & m_{43} \\ m_{14} & m_{24} & m_{34} & m_{44} \\  \end{pmatrix}  = \begin{pmatrix}
        0 & 1 & 0 & 0 \\ -1  & 0 & 0 & 0 \\ 0 & 0 & 0 & 1 \\ 0 & 0 & -1 & 0
    \end{pmatrix}.
    \end{align*}
\end{widetext}
This yields 5 independent non-linear equations (after accounting for redundancy in the original 16). 

Yet, this is not enough equations to solve for the remaining 8 unknown transfer matrix elements. Just as we measured the centroids of the 4D phase space to obtain information about the transfer matrix, we can also measure higher-order statistical moments of the beam. The next order statistical moment is the beam covariance matrix, which evolves in the beamline with the transfer matrix by,
\begin{widetext}
\begin{align}
\label{eq:beamcovpropagation}
       \mathbf{M^{-1}} \begin{pmatrix}
        \left< xx\right> & \left<xp_x \right> & \left<xy \right> & \left<xp_y \right> 
        \\
        \left<xp_x \right> & \left<p_xp_x \right> & \left<yp_x \right> & \left<p_xp_y \right> 
        \\
        \left<xy \right> & \left<yp_x \right> & \left<yy \right> & \left<yp_y \right> 
        \\
        \left<xp_y \right> & \left<p_xp_y \right> & \left<yp_y \right> & \left<p_yp_y \right> 
    \end{pmatrix}_f \mathbf{M^{-1 T}} =    \begin{pmatrix} \left< xx\right> & \left<xp_x \right> & \left<xy \right> & \left<xp_y \right> 
        \\
        \left<xp_x \right> & \left<p_xp_x \right> & \left<yp_x \right> & \left<p_xp_y \right> 
        \\
        \left<xy \right> & \left<yp_x \right> & \left<yy \right> & \left<yp_y \right> 
        \\
        \left<xp_y \right> & \left<p_xp_y \right> & \left<yp_y \right> & \left<p_yp_y \right> 
    \end{pmatrix}_i,
\end{align}
\end{widetext}
where the final beam covariance matrix on the left-hand side is measured. Since we choose to use a cathode that is disordered and polycrystalline, as is typical, we assume that the position and momentum are independent, or in other words, the source position-momentum correlation is zero. This yields another 4 independent non-linear equations, $\left < x p_x\right>_i = \left < x p_y\right>_i = \left < y p_x\right>_i = \left < y p_y\right>_i = 0$, which is related to the measured beam covariances and transfer matrix by Eq. \ref{eq:beamcovpropagation}. Note that this average is taken over the whole beam, and nonzero correlation is permitted locally. As a result, this provides a total of 9 independent non-linear equations. 

There are 8 unknown transfer matrix elements, fewer than the number of equations, making this an overdetermined system of equations. We can then apply the Gauss-Newton method \cite{supp} to numerically find a best-fit solution, which yields the full transfer matrix. Once the full transfer matrix is acquired, it can be inverted and applied to a measured 4D phase space to reconstruct the entire 4D phase space at emission. Our procedure, via the symplectic condition, enforces unity determinant of the transfer matrix and preserves 4D emittance; since we are using physical momenta, this remains true in transfer matrices with acceleration.

\section{Experimental Demonstration Methods}
\label{sec:exp_setup}

\begin{figure*}
    \centering
    \includegraphics[width = 1\linewidth]{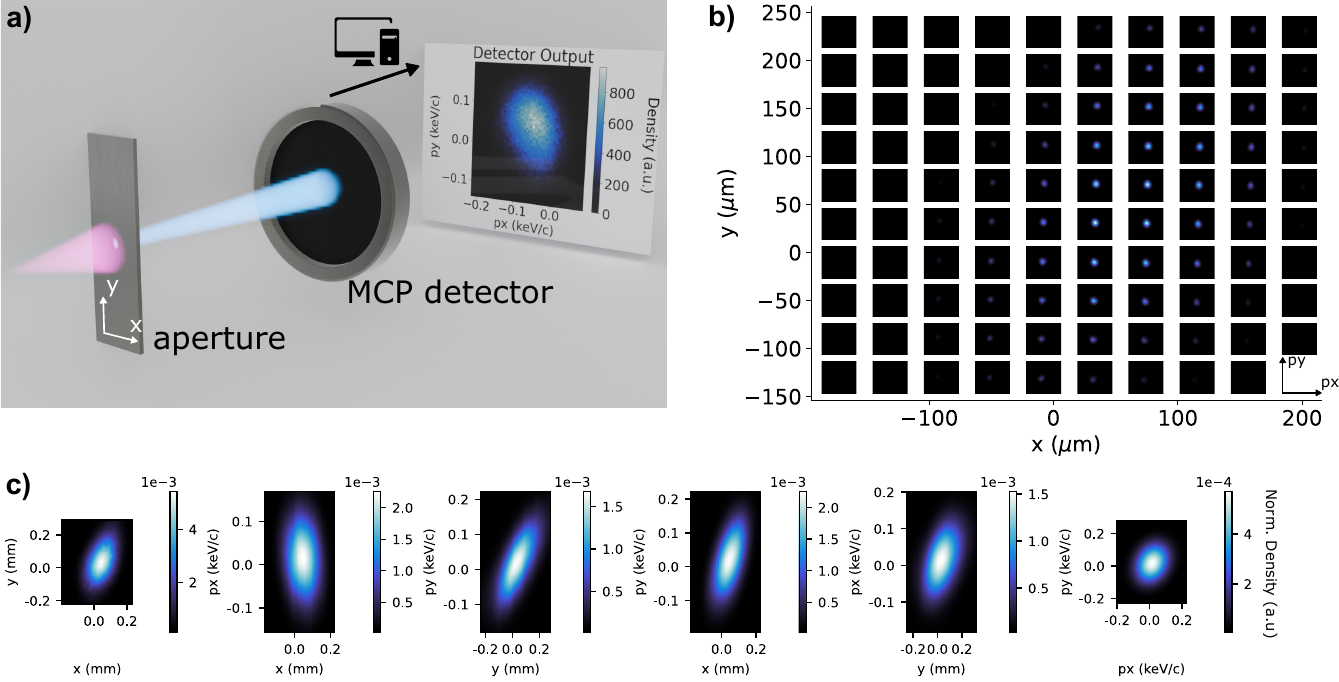}
    \caption{\textbf{a)} Diagram of the aperture scan measurement. Beam (pink) is scanned over the aperture, and the transverse spatial distribution of the transmitted beam (blue) is imaged by the detector, which is converted into a transverse momentum distribution. \textbf{b)} Example of a measured 4D phase space, where a subset of the data is shown. Each image is a $(p_x,p_y)$ distribution measured when the beam is at the $(x,y)$ position on the aperture. \textbf{c)} The 4D phase space in \textbf{b)} projected into 2D planes along the $(x,p_x,y,p_y)$ axes.}
    \label{fig:beamline}
\end{figure*}
 
We measured the 4D phase space of a beam using the PHOEBE (PHOtocathode Epitaxy and Beam Experiments) test beamline at Cornell University \cite{galdi:ipac2024-mopr78}. Photoemission is driven at the cathode by a continuous-wave diode laser, and the position of the laser on the cathode is controlled by mirrors mounted on motorized linear translation stages. A beamsplitter splits the laser right before it enters the gun chamber, and the reflected laser is viewed by a CMOS camera acting as a virtual cathode, which measures the laser position. 

For the purposes of demonstrating this method, we used a micro-patterned alkali-antimonide cathode - see the Supplemental Materials for more details \cite{supp}. The cathode consisted of a Si/SiO$_2$ substrate with a layer of copper deposited on top with a periodic lattice of fiducial markers of varying geometries, such that the underlying Si/SiO$_2$ is exposed. A thin film of Na-K-Sb was then grown on top of the substrate using molecular beam epitaxy. The quantum efficiency of the Na-K-Sb film grown on the copper is larger than Si/SiO$_2$; therefore periodically modulating the quantum efficiency across the cathode when photoemission is driven at 405 nm.

Electrons emitted by the cathode are accelerated by a DC gun biased at 15 kV. The beam is focused by a solenoid onto a circular aperture with a diameter of 30 $\mu$m at which the 4D phase space of the beam is measured, as shown in Fig \ref{fig:beamline}a. After the aperture, the transmitted beam traveled through a $\sim$0.5 m drift section until it reached the detector. The detector consisted of a microchannel plate, which multiplied electrons and accelerated them onto a YAG scintillator screen imaged by a sCMOS camera. 

Approximating the aperture as a point source, the transverse spatial distribution imaged by the detector trivially yields the transverse momentum distribution of the transmitted beam by dividing the detector position by the drift length and multiplying by the longitudinal momentum, which is inferred from the DC gun voltage bias. This transverse momentum distribution was measured continuously as a dipole magnet scans the beam across the aperture in a 2D grid. Each step of the 2D grid of dipole magnet currents corresponds to a point in the $x$-$y$ position phase space. As such, for each point in the $x$-$y$ position phase space, there is an associated transverse momentum distribution. After subtracting the momentum kick added by the dipole magnet from the measured transverse momentum, the 4D phase space can be composed, an example of which is shown in Fig. \ref{fig:beamline}b. For analysis, we will visualize the 4D phase space in 2D projections onto the transverse coordinate axes, like in Fig. \ref{fig:beamline}c. 

To measure the partial transfer matrix, we use a small laser spot size such that the beam is solely emitted from the uniform regions of the cathode, excluding the patterned fiducial markers. The laser is moved around the cathode, within a 200 $\mu$m by 200 $\mu$m region, with the 4D phase space of the beam measured at 8 positions. The 4D phase spaces measured at each laser position is similar to that shown in Fig. \ref{fig:beamline}b-c. The clean phase spaces allow us to fit the 2D phase space projections (e.g. Fig. \ref{fig:beamline}c) to a 2D supergaussian \cite{supp}, which then supplies the beam covariances required to enforce zero source position-momentum correlation when solving for the full 4D transfer matrix. We then switch to a larger laser spot that covers the patterned fiducial markers, measure the 4D phase space of the resulting beam, and reconstruct the source 4D phase space using the solved transfer matrix.

\section{Experimental Results}
\label{sec:results}
\begin{figure*}
    \centering
    \includegraphics[width=1\linewidth]{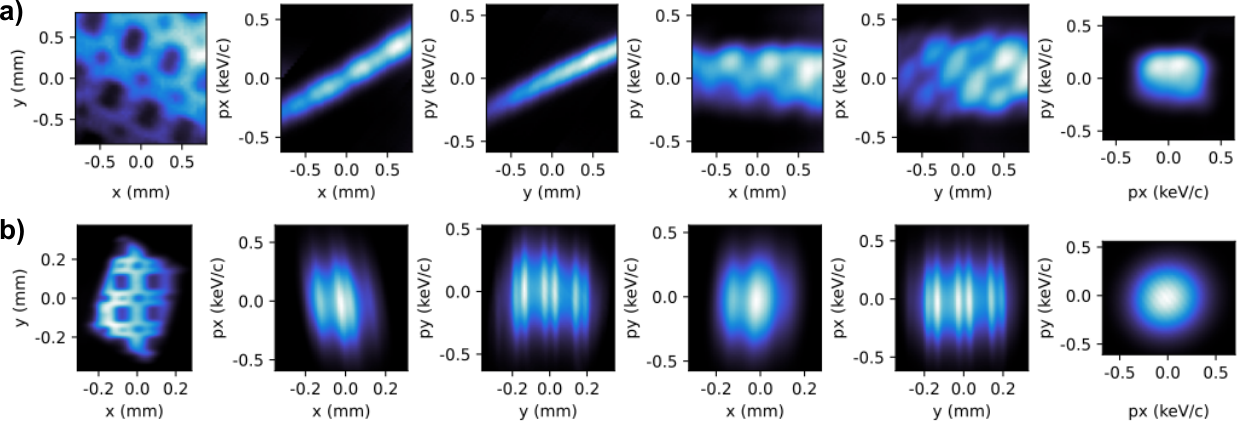}
    \caption{\textbf{a)} Measured 4D phase space of the beam at the aperture. Only a 1200 $\mu$m by 1200 $\mu$m part of the beam was measured. \textbf{b)} Reconstructed source 4D phase space of the beam at the cathode.}
    \label{fig:full_reconstruction}
\end{figure*}
Driving the photoemission with a large laser spot that covers multiple patterned markers on the cathode, we measure a 4D phase space using the aperture scan, shown in Fig. \ref{fig:full_reconstruction}a. During the aperture scan, we only scan over a part of the entire beam in the $x$-$y$ position phase space to increase the 2D spatial resolution. Both the hard edges of the scan range and the modulation of quantum efficiency on the cathode surface lead to sharp edges and intricate intensity modulations not present in the example phase space shown in Fig. \ref{fig:beamline}c. Following the process laid out in Sec. \ref{sec:theory},  we reconstruct the source 4D phase space of the beam at the cathode, shown in Fig. \ref{fig:full_reconstruction}b.


Upon first inspection of the reconstructed phase space in Fig. \ref{fig:full_reconstruction}b, we note that the position-momentum phase spaces ($x$-$p_x$, $y$-$p_y$, $x$-$p_y$, $y$-$p_x$) display little correlation between the position and momentum, on average over all the electrons, which is expected from a randomly disordered cathode. Since we do not use the measured 4D phase space (Fig. \ref{fig:full_reconstruction}a) in solving for the transfer matrix, this is not trivially a result of constraining the source position-momentum correlation to be zero. 

The $p_x$-$p_y$ momentum phase space also displays little average correlation and is highly symmetric, which is expected for a polycrystalline cathode where the band structure is not well-defined and has a randomly disordered surface that allows electrons to uniformly scatter into all energetically allowed transverse momentum \cite{saha2023theory,sid_surfacedisorder}. The fact that this simple momentum distribution arises from such a complex phase space, without enforcing it as a constraint on the transfer matrix, is evidence that the reconstruction does indeed produce physically sensible results.
 
\begin{figure*}
    \centering
    \includegraphics[width=1\linewidth]{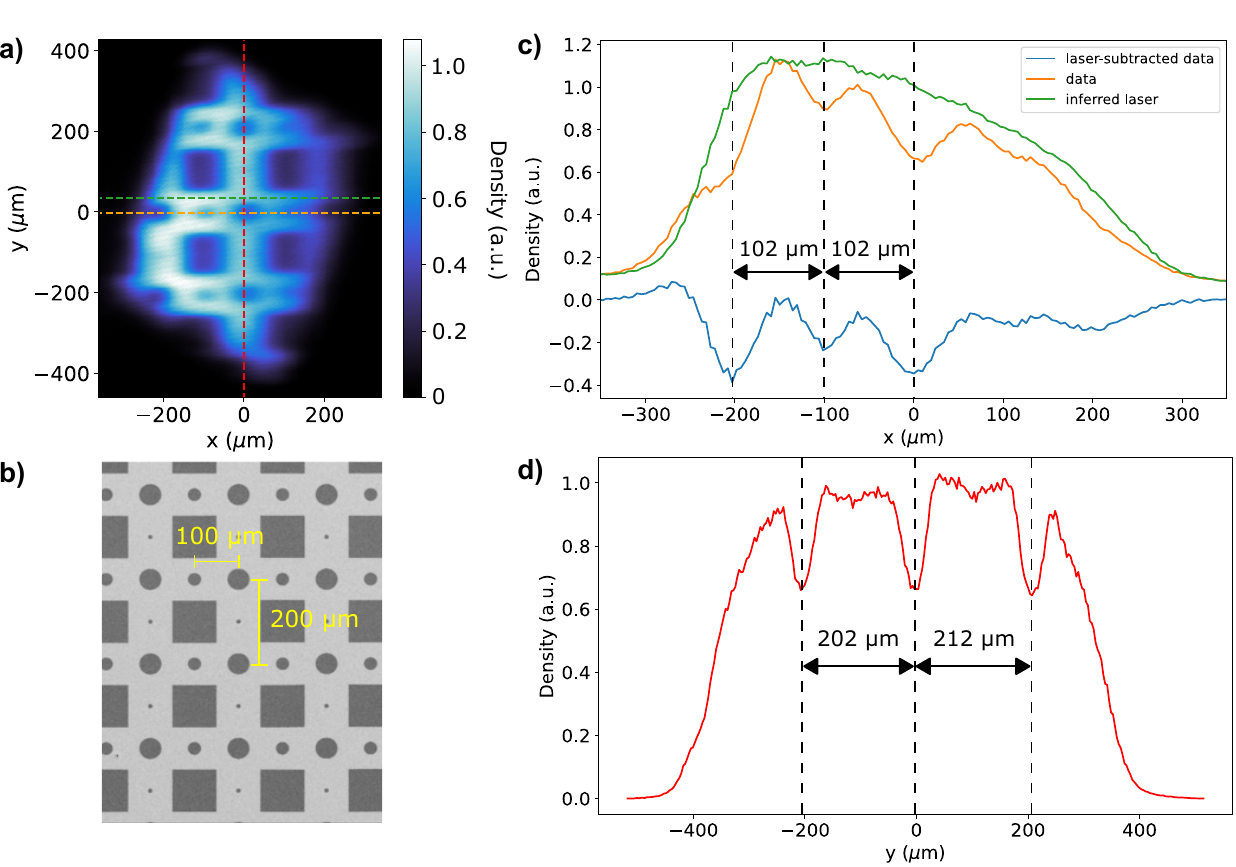}
    \caption{\textbf{a)} Detailed image of the reconstructed source $x$-$y$ position phase space. Linecut profiles are indicated by the dashed lines and correspondingly colored with the displayed profiles in \textbf{c)} and \textbf{d)}. \textbf{b)} Scanning electron microscope image of the substrate. Dark regions correspond to SiO$_2$, and copper for light regions. \textbf{c)} Horizontal linecut profiles, taken on (orange) and off (green) the features in \textbf{a)}. The off-feature linecut is treated as background and subtracted from the on-feature linecut, yielding the linecut in blue. \textbf{d)} Vertical linecut profile. Black dashed lines in \textbf{c)} and \textbf{d)} indicate the position of features corresponding to known QE modulations induced by the substrate pattern.}
    \label{fig:dot_analysis}
\end{figure*}

We analyze the reconstructed $x$-$y$ position phase space, shown in Fig. \ref{fig:dot_analysis}a, by investigating linecut profiles of the density taken across features induced by the substrate micro-pattern, shown in Fig. \ref{fig:dot_analysis}b. Looking at the horizontal linecut profile in Fig. \ref{fig:dot_analysis}c, there is a noticeable background slope in the data, which can be attributed to variation in the laser intensity. To account for this, we take a linecut profile nearby, which does not intersect the substrate micro-pattern features, and subtract it from the data. The features appear as troughs in the density, and we measure the center-to-center distance. The average center-to-center distance comes out to $102 \pm 4$ $\mu$m, which matches the spacing between the corresponding substrate fiducial markers, measured by a scanning electron microscope (SEM) in Fig. \ref{fig:dot_analysis}b to be 100 $\mu$m. The uncertainties of the average center-to-center distances here are propagated from the resolution of the aperture scan measurement.  Similarly, we measure the center-to-center distance between features in a vertical linecut profile in Fig. \ref{fig:dot_analysis}d, and the average center-to-center distance comes out to $207 \pm 5$ $\mu$m. This is slightly larger than the 200 $\mu$m spacing measured by an SEM, but still within $2\sigma$. While acknowledging that some deviations may occur due to the cathode growth and QE modulations not perfectly corresponding with the substrate micro-pattern, we nevertheless observe relatively good agreement between the position of features in the reconstructed $x$-$y$ position phase space and the position of the corresponding fiducial markers measured by SEM.

As noted earlier, we only measured part of the beam in Fig. \ref{fig:full_reconstruction}a, so reconstructions of the source 4D phase space could be frustrated since we may not be measuring all of the electrons being emitted from the reconstructed region on the cathode. To understand this effect, we performed General Particle Tracer simulations, tracking the beam emitted from a point on the cathode with the measured MTE to the aperture. From these simulations, we found that the beam forms a roughly $45$ $\mu$m rms spot at the aperture. Given that we scan the beam within a $1200$ $\mu$m by $1200$ $\mu$m region, we can safely say that we measure all of the electrons emitted from the region of the cathode that was reconstructed in the $x$-$y$ position phase space of  Fig. \ref{fig:full_reconstruction}b except for near the edges of that phase space.
\begin{figure*}
    \centering
    \includegraphics[width=1\linewidth]{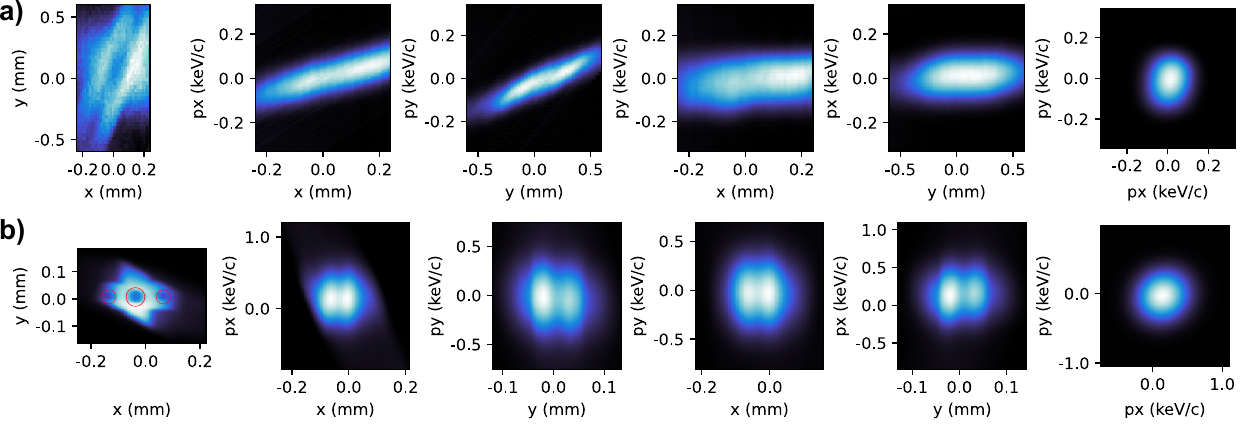}
    \caption{\textbf{a)} 2D projections of a 4D phase space of the beam measured at beamline optics settings where the intensity modulations in the $x$-$y$ phase space are significantly distorted. \textbf{b)} Source 4D phase space reconstructed from \textbf{a)}.  In the source spatial distribution, the dark circular features within the red circles correspond with fiducial markers on the cathode.}
    \label{fig:crazy_dots}
\end{figure*}

We also attempt to reconstruct the source 4D phase space of the beam from a 4D phase space, shown in Fig. \ref{fig:crazy_dots}a, measured by performing the aperture scan at a different solenoid focusing strength. The intensity modulation in the $x$-$y$ position phase space here (Fig. \ref{fig:crazy_dots}a) is now significantly distorted compared to its counterpart in the 4D phase space that we analyzed earlier (Fig. \ref{fig:full_reconstruction}a). In the reconstructed source $x$-$y$ position phase space, shown in Fig. \ref{fig:crazy_dots}b, the average center-to-center distance between the circular features is $94 \pm 6$ $\mu$m. This agrees with the distance between the corresponding fiducial markers on the cathode (Fig. \ref{fig:dot_analysis}b), measured by SEM imaging to be 100 $\mu$m. The accurate reconstruction of the source spatial distribution from a distorted 4D phase space signifies that our calculated transfer matrix successfully incorporates the momentum of the measured beam to ``unscramble" the distorted $x$-$y$ phase space that was measured at the aperture.

\begin{figure}
    \centering
    \includegraphics[width=1\linewidth]{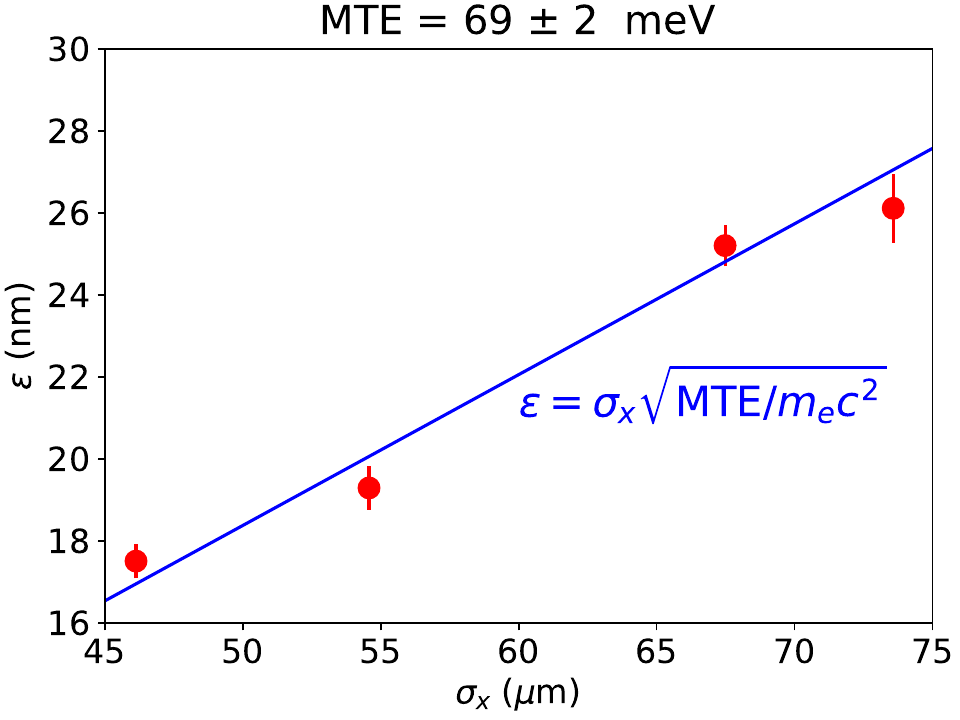}
    \caption{4D emittance, $\epsilon$, measured as a function of beam size at the cathode, $\sigma_x$, and linearly fitted to yield an MTE of $69 \pm 2$ meV. The fit requires $\epsilon \rightarrow 0$ as $\sigma_x \rightarrow 0$.}
    \label{fig:emitvsspotsize}
\end{figure}

The other primary goal of reconstructing source 4D phase space is to accurately measure the MTE of the cathode. To check that the MTE calculated from the reconstructed source momentum distribution is correct, we compare it to the MTE measured by an emittance measurement via the aperture scan. 
We measured 4D phase spaces while varying the source beam sizes at the cathode, $\sigma_x$, and reconstructed the source 4D phase spaces for each source beam size. Fitting each reconstructed source $p_x$-$p_y$ momentum phase space to a 2D supergaussian \cite{supp} for $\left< p_xp_x\right>_i$, $\left< p_yp_y\right>_i$, and $\left< p_x p_y\right>_i$, we calculate the MTE as given by Sec. \ref{sec:theory} to be $73 \pm 2$ meV. The uncertainty of the MTE here is propagated from the uncertainty of the measured partial transfer matrix. To check this MTE value, we calculated the 4D emittance \cite{supp}, $\epsilon$, of each measured 4D phase space and linearly fitted it as a function of $\sigma_x$, $\epsilon = \sigma_x\sqrt{\text{MTE}/m_ec^2}$, where $m_e$ is the electron mass. This is shown in Fig. \ref{fig:emitvsspotsize}, yielding a MTE of $69 \pm 2$ meV. This fitted MTE value agrees with the MTE calculated from the reconstructed $p_x$-$p_y$ momentum phase spaces, validating the source $p_x$-$p_y$ momentum phase space reconstruction. The MTE we expect for a NaKSb cathode can be calculated given the photo-excitation energy, $h \nu \approx 3$ eV, and the work function, $\phi \approx 2$ eV \cite{spicer1958photoemissive,cultrera2013growth}, as $\text{MTE} = (h \nu - \phi)/3 \approx 330$ meV \cite{DS-theory}. In comparison, the MTE we measure is quite small, but this is not necessarily unexpected given that the cathode quantum efficiency has degraded by a few orders of magnitude since it was grown, and we are likely emitting near the photoemission threshold.

The agreement between the position of features measured in the reconstructed source position phase space and the known position of fiducial markers in the substrate micro-pattern suggests that we have correctly reconstructed the source position phase space of the beam. Together with the confirmation of the MTE and the low position-momentum correlations aligning with our expectations of the cathode's electronic and physical structure, we have likely reconstructed the entire source 4D phase space with good fidelity. 

The reconstructed source 4D phase space still includes measurement errors and artifacts such as background noise, point spread function blurring of the measured phase space, and others. We now discuss these measurement errors and artifacts in detail and how they are accounted for in our prior analysis. 

\section{Error Considerations}
\label{sec:error}
\subsection{Camera Background Noise}
\begin{figure*}
    \centering
    \includegraphics[width=1\linewidth]{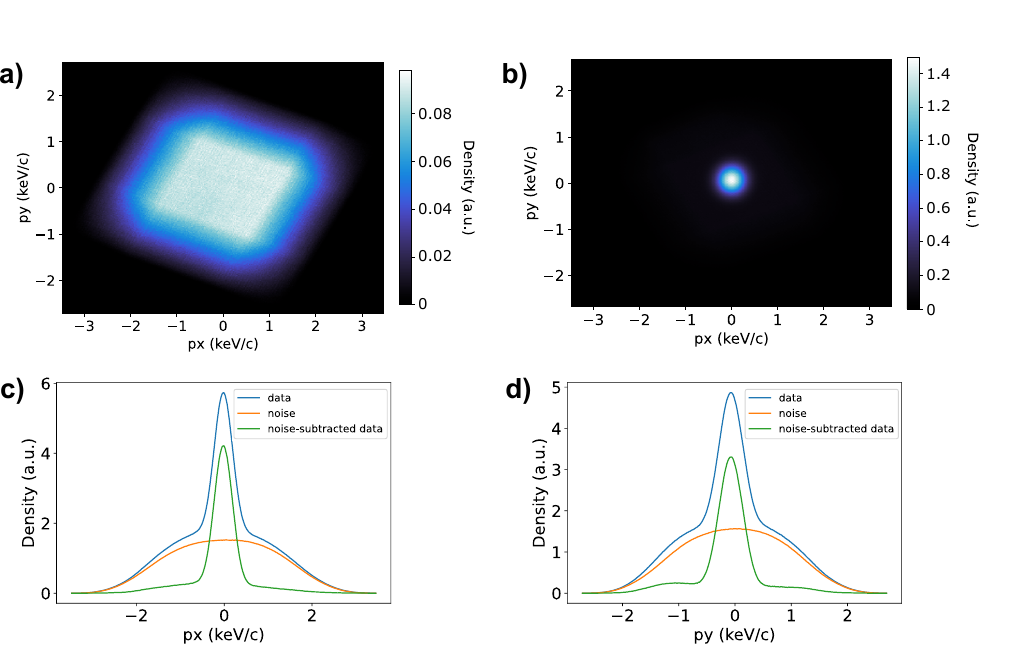}
    
    \caption{\textbf{a)} Isolated background noise in reconstructed $p_x$-$p_y$ momentum space due to camera background noise. \textbf{b)} Measured source $p_x$-$p_y$ momentum space of a beam. \textbf{c)} and \textbf{d)} show, respectively, projections of the background noise (green) and the measured source $p_x$-$p_y$ momentum space (blue) onto the $p_x$ and $p_y$ axes, as well as the background-subtracted distribution (orange).}
    \label{fig:noise_analysis}
\end{figure*}

While measuring the 4D phase space, there is an intrinsic background noise added by the detector itself. To account for this, we perform an aperture scan without a beam, achieved by blocking the driving laser and turning off the DC gun power supply, which yields the background noise in the measured 4D phase spaces. Having solved for the full transfer matrix in previous measurements, we can invert it and apply it to the measured 4D background noise, yielding the background noise in the source 4D phase space. This background noise can then be subtracted from the reconstructed source 4D phase space of a beam, the resulting difference being shown earlier in Fig. \ref{fig:full_reconstruction}b. 

Here we focus on the reconstructed $p_x$-$p_y$ momentum phase space, shown in Fig. \ref{fig:noise_analysis}a, but the full source 4D phase space and its background noise can be found in the Supplemental Materials \cite{supp}. Projecting the reconstructed source $p_x$-$p_y$ background noise onto either $p_x$ and $p_y$, as in Fig. \ref{fig:noise_analysis}c and \ref{fig:noise_analysis}d, we can see that the background noise appears Gaussian-like. Doing the same 1D projection for the reconstructed $p_x$-$p_y$ momentum phase space of the beam we measured previously, fully shown here in Fig. \ref{fig:noise_analysis}b, we see that the background noise matches the tail of the data distribution. Subtracting the background noise from the data then produces a much cleaner distribution that can be easily fitted with a supergaussian and yields a more accurate MTE. 

There is some residual background leftover after the background noise subtraction, which can be attributed to other sources of noise, such as the presence of stray beam leaking out past the aperture sidewalls and hitting the detector. Evidence of stray beam can be seen 
in the phase space fit residuals, shown in Fig. S2b in the Supplemental Materials \cite{supp}.

\subsection{Effect of finite aperture size and detector point spread function}

When measuring the 4D phase space of the beam with the aperture scan, the finite size of the aperture will blur the measured $x$-$y$ position phase space. To understand this, we can imagine the case where the beam is a point source. Scanning the beam across the aperture then produces an $x$-$y$ position phase space of uniform density that matches the shape of the aperture, a circle with diameter $A$. In the case of a beam of non-zero size at the aperture, the measured $x$-$y$ position phase space is then a convolution of the aperture with the actual $x$-$y$ position phase space, where, for simplicity, we represent the aperture as a Gaussian with an rms spread of  $\sigma_A = A/\sqrt{12}$. As a result, the finite aperture size blurs the $x$-$y$ position phase space.

The effect of the detector point spread function (PSF) is similar, where a point source object is imaged by the detector as a Gaussian with an rms spread of $\sigma_{D}$. Like the $x$-$y$ position phase space, the detector PSF is convolved with the actual $p_x$-$p_y$ momentum phase space, thus blurring it as well. Since it is so similar, we will also use the term PSF to refer to both aperture and detector blurring.

Knowing the detector PSF, we can deconvolve the detector images of the beam measured during the aperture scan to obtain the true 4D phase space. However, direct deconvolution can produce inaccurate results in the case where the signal-to-noise ratio is low. Instead, if we are only interested in statistical quantities like MTE, we need only to account for the aperture and detector PSFs in the measured beam covariance matrix and propagate it to the beam covariance at the source.

Assuming the $x$-$y$ and $p_x$-$p_y$ phase spaces are 2D Gaussians, the convolution of the phase space with the aperture and detector PSFs results in the following relationship between the actual beam covariance matrix $\Sigma_f'$ and the measured beam covariance matrix $\Sigma_f$,
\begin{align*}
       \Sigma_f' = \Sigma_f - \begin{pmatrix}
        \sigma_A^2&  & & & 
        \\
        & \sigma_D^2 & &
        \\
        & & \sigma_A^2 &
        \\
        & & & \sigma_D^2
    \end{pmatrix}
    \end{align*}
We can then propagate the PSF-corrected beam covariances to the cathode by applying the inverted transfer matrix to $\Sigma_f'$, like in Eq. \ref{eq:beamcovpropagation}, yielding the source beam covariance matrix corrected for the aperture and detector PSF. For measuring MTE, we are particularly interested in the corrected source momentum covariances, $\left<p_xp_x\right>_i'$, $\left<p_yp_y\right>_i'$, and $\left<p_xp_y\right>_i'$,
\begin{align*}
    \left<p_xp_x\right>_i' = & \left<p_xp_x\right>_i - m_{11}^2 \sigma_{D}^2 - m_{21}^2 \sigma_{A}^2 - m_{31}^2\sigma_{D}^2  \\& -  m_{41}^2\sigma_{A}^2
    \\
     \left<p_yp_y\right>_i' = & \left<p_yp_y\right>_i -  m_{13}^2\sigma_{D}^2 - m_{23}^2\sigma_{A}^2  -  m_{33}^2\sigma_{D}^2 \\ & - m_{43}^2\sigma_{A}^2
     \\
     \left<p_xp_y\right>_i'  = & \left<p_xp_y\right>_i -  m_{11}m_{13}\sigma_{D}^2 - m_{21} m_{23}\sigma_{A}^2 
     \\
     &  -  m_{31}m_{33}\sigma_{D}^2 -m_{41}m_{43}\sigma_{A}^2 
\end{align*}
Here, $\left<p_xp_x\right>_i$, $\left<p_yp_y\right>_i$, and $\left<p_xp_y\right>_i$ are calculated from fitting the reconstructed source momentum phase space. With our measured source momentum phase space, this results in a roughly 15\% decrease in the MTE when using an aperture with a diameter of 30 $\mu$m and a detector with a PSF spread of $\sigma_D = 100$ $\mu$m, which yields the aforementioned MTE of $73 \pm 2$ meV in Sec. \ref{sec:results}. 

\subsection{Other sources of error}

Experimental drift in the beam phase space over time may cause subsequent measurements of the beamline transfer matrix and source phase space to be inaccurate. This drift may originate from small changes in the laser optics positions due to vibrations or temperature shifts, time-varying laser intensities, and the cathode QE and MTE changing over time. To address changes in the laser position and intensity, we monitored them using the virtual cathode camera over the course of an aperture scan measurement. We did not observe any significant drift in the laser position. However, there are noticeable variations in the laser intensity. Assuming linear photoemission, we account for this by normalizing the images of the beam taken during the aperture scan by the laser intensity measured simultaneously. 
As for cathode QE and MTE, we did not observe significant changes in either over the typical timescale of a measurement, which is roughly 15 minutes for an aperture scan. We performed multiple aperture scans over 12 hours and saw no significant drift in the beam emittance calculated from the measured phase spaces. 

\section{Conclusions}

In this paper, we present a method for determining the 4D transfer matrix of a beamline by using a simple aperture scan diagnostic to measure the dependence of the 4D phase space on the source emission position. This method is experimentally applied to find the transfer matrix of the PHOEBE test beamline and reconstruct the source 4D phase space of a beam emitted from a micro-patterned cathode. Analyzing the reconstructed source 4D phase spaces, we fit the $p_x$-$p_y$ momentum phase space to directly determine the MTE and found that it agreed with aperture scan measurements of MTE. We also confirmed that the source $x$-$y$ position phase space is reconstructed successfully across varying beamline optics settings by observing that the spacing between features in the reconstructed $x$-$y$ position phase space matches that of the corresponding fiducial markers patterned onto the cathode. Altogether, this is suggestive that we are able to solve for the beam transfer matrix and apply it to reconstruct the source 4D phase space with good fidelity.

A feature of this method is the ability to image source position-momentum correlation at microscopic scales. We could exploit this ability in future work by deterministically engineering physical features, like steps in height on the cathode surface, or chemical features, such as interfaces between deposited metals of different work functions, and studying the effects on the momentum of electrons emitted nearby. Studies like these, aiming to understand local position-momentum correlations, could open up the development of ``designer photocathodes" with physical and chemical surface features engineered to shape the emission of electrons in a controllable fashion. 

Finally, this source reconstruction method is accessible to existing accelerators, requiring only linear beam transport, a method of measuring 4D phase space (aperture scans, pepperpot, etc), and a means to deterministically control the source position (or momentum) of the beam. By requiring beam transport to be linear, accelerators must operate at beam currents where space charge is negligible and include only linear optics. In addition, there can be no coupling of the transverse coordinates to time or energy, as they are not measured by 4D phase space measurements. However, this method could be extended to include time and energy in the transfer matrix and source phase space if the 5D or 6D phase space of the beam is measured. Given that all these conditions are satisfied, our source reconstruction method can be applied to measure the beamline transfer matrix for arbitrary linear optics and the phase space of the beam at the source.  


\section{Acknowledgments}
This work was supported by U.S. Department of Energy, Grants DE-SC0020144 and DE-AC02-76SF00515, and U.S. National Science Foundation Grant No. PHY-1549132, the Center for Bright Beams.

The authors would also like to thank the CLASSE technical staff for their assistance with the building and maintenance of the PHOEBE test beamline and colleagues within the Center for Bright Beams for many insightful discussions.

\bibliography{apssamp.bib}

\end{document}



\title{\textbf{Supplement: Experimental Reconstruction of 4D Phase Space at the Cathode} 
}%

\author{Charles Zhang $^1$, Elena Echeverria$^1$, Abigail Flint$^1$, William Li $^2$, Christopher M. Pierce $^3$, Alice Galdi $^4$, Chad Pennington $^5$, Adam Bartnik$^1$, Ivan Bazarov$^1$, Jared Maxson$^1$}

\affiliation{$^1$Cornell Laboratory for Accelerator-Based Sciences and Education, Cornell University, Ithaca, NY, USA}
\affiliation{$^2$Brookhaven National Laboratory, Upton, NY, USA}
\affiliation{$^3$Enrico Fermi Institute, University of Chicago, Chicago, IL 60637, USA}
\affiliation{$^4$Department of Industrial Engineering, University of Salerno, Fisciano (SA) Italy}
\affiliation{$^5$University of California, Los Angeles, CA, USA}
\date{\today}


\maketitle

\section{Gauss-Newton Method for Solving 4D Transfer Matrix}

We construct a system of equations involving the 4D transfer matrix and solve for the transfer matrix by applying the Gauss-Newton method, a commonly used numerical technique for minimizing the sum of squares of functions representing the equations.

From expanding the symplectic condition, $\mathbf{M}\Omega\mathbf{M}^T = \Omega$,
\begin{align*}
        \begin{pmatrix} m_{11} & m_{12} & m_{13} & m_{14} \\ m_{21} & m_{22} & m_{23} & m_{24} \\ m_{31} & m_{32} & m_{33} & m_{34} \\ m_{41} & m_{42} & m_{43} & m_{44} \\  \end{pmatrix}  \begin{pmatrix}
        0 & 1 & 0 & 0 \\ -1  & 0 & 0 & 0 \\ 0 & 0 & 0 & 1 \\ 0 & 0 & -1 & 0
    \end{pmatrix}  \begin{pmatrix} m_{11} & m_{21} & m_{31} & m_{41} \\ m_{12} & m_{22} & m_{32} & m_{42} \\ m_{13} & m_{23} & m_{33} & m_{43} \\ m_{14} & m_{24} & m_{34} & m_{44} \\  \end{pmatrix}  = \begin{pmatrix}
        0 & 1 & 0 & 0 \\ -1  & 0 & 0 & 0 \\ 0 & 0 & 0 & 1 \\ 0 & 0 & -1 & 0
    \end{pmatrix}, 
    \end{align*}
we acquire the following system of 5 non-linear equations that constrains the transfer matrix,
\begin{align*}
    -1 - m_{12}  m_{21} + m_{11}  m_{22} - m_{32}  m_{41} + m_{31}  m_{42},  -m_{14}  m_{21} & = 0
    \\
 m_{11}  m_{24} - m_{34}  m_{41} + m_{31}  m_{44}  &= 0
 \\
 -m_{13}  m_{22} + m_{12}  m_{23} - m_{33}  m_{42} + m_{32}  m_{43} & = 0
 \\
 -1 - m_{14}  m_{23} + m_{13}  m_{24 }- m_{34}  m_{43} + m_{33}  m_{44} & = 0
 \\
 -m_{14} m_{22} + m_{12} m_{24} - m_{34} m_{42} + m_{32} m_{44} & = 0
\end{align*}
We also use the relationship between the source beam covariance matrix and the beam covariance measured by an aperture scan, 
\begin{align}
\label{eq:beamcovpropagation}
       \mathbf{M^{-1}} \begin{pmatrix}
        \left< xx\right> & \left<xp_x \right> & \left<xy \right> & \left<xp_y \right> 
        \\
        \left<xp_x \right> & \left<p_xp_x \right> & \left<yp_x \right> & \left<p_xp_y \right> 
        \\
        \left<xy \right> & \left<yp_x \right> & \left<yy \right> & \left<yp_y \right> 
        \\
        \left<xp_y \right> & \left<p_xp_y \right> & \left<yp_y \right> & \left<p_yp_y \right> 
    \end{pmatrix}_f \left(\mathbf{M^{-1}}\right)^T =    \begin{pmatrix} \left< xx\right> & \left<xp_x \right> & \left<xy \right> & \left<xp_y \right> 
        \\
        \left<xp_x \right> & \left<p_xp_x \right> & \left<yp_x \right> & \left<p_xp_y \right> 
        \\
        \left<xy \right> & \left<yp_x \right> & \left<yy \right> & \left<yp_y \right> 
        \\
        \left<xp_y \right> & \left<p_xp_y \right> & \left<yp_y \right> & \left<p_yp_y \right> 
    \end{pmatrix}_i,
\end{align}
Assuming that the position and momentum are uncorrelated at the cathode, $\left < x p_x\right>_i = \left < x p_y\right>_i = \left < y p_x\right>_i = \left < y p_y\right>_i = 0$, and noting that the inverse transfer matrix, $\mathbf{M^{-1}}$, can be expressed through the symplectic condition as,
\begin{align}
    \mathbf{M^{-1}}
    =  -\Omega \mathbf{M^T} \Omega = 
    \begin{pmatrix}
m_{22} && -m_{12} && m_{42} && -m_{32}\\
-m_{21} && m_{11} && -m_{41} && m_{31}\\
m_{24} && -m_{14} && m_{44} && -m_{34}\\
-m_{23} && m_{13} && -m_{43} && m_{33}\\
\end{pmatrix},
\end{align}
we obtain 4 additional equations,
\begin{align*}
  &  -\left< p_x p_x \right>_f m_{11} m_{12} + \left< x p_x \right>_f m_{12} m_{21} + \left< x p_x \right>_f m_{11} m_{22} - \left< x x \right>_f m_{21} m_{22} - 
 \left< p_x p_y \right>_f m_{12} m_{31}  \\ & +  \left< x p_y \right>_f m_{22} m_{31} - \left< p_x p_y \right>_f m_{11} m_{32} + \left< x p_y \right>_f m_{21} m_{32} - 
 \left< p_y p_y \right>_f m_{31} m_{32} + \left< y p_x \right>_f m_{12} m_{41}  \\ & - \left< x y \right>_f m_{22} m_{41} + \left< y p_y \right>_f m_{32} m_{41} + 
 \left< y p_x \right>_f m_{11} m_{42} - \left< x y \right>_f m_{21} m_{42} + \left< y p_y \right>_f m_{31} m_{42}\\ & - \left< y y \right>_f m_{41} m_{42}  = 0
 \\\\
& -\left< p_x p_x \right>_f m_{12} m_{13} + \left< x p_x \right>_f m_{13} m_{22} + \left< x p_x \right>_f m_{12} m_{23} - \left< x x \right>_f m_{22} m_{23} - 
 \left< p_x p_y \right>_f m_{13} m_{32} \\ & + \left< x p_y \right>_f m_{23} m_{32} - \left< p_x p_y \right>_f m_{12} m_{33} + \left< x p_y \right>_f m_{22} m_{33} - 
 \left< p_y p_y \right>_f m_{32} m_{33} + \left< y p_x \right>_f m_{13} m_{42}  \\ & - \left< x y \right>_f m_{23} m_{42} + \left< y p_y \right>_f m_{33} m_{42} + 
 \left< y p_x \right>_f m_{12} m_{43} - \left< x y \right>_f m_{22} m_{43} + \left< y p_y \right>_f m_{32} m_{43} \\ & - \left< y y \right>_f m_{42} m_{43} =0
 \\\\
& -\left< p_x p_x \right>_f m_{11} m_{14} + \left< x p_x \right>_f m_{14} m_{21} + \left< x p_x \right>_f m_{11} m_{24} - \left< x x \right>_f m_{21} m_{24} - 
 \left< p_x p_y \right>_f m_{14} m_{31} \\ & + \left< x p_y \right>_f m_{24} m_{31} - \left< p_x p_y \right>_f m_{11} m_{34} + \left< x p_y \right>_f m_{21} m_{34} - 
 \left< p_y p_y \right>_f m_{31} m_{34} + \left< y p_x \right>_f m_{14} m_{41}  \\ & - \left< x y \right>_f m_{24} m_{41} + \left< y p_y \right>_f m_{34} m_{41} + 
 \left< y p_x \right>_f m_{11} m_{44} - \left< x y \right>_f m_{21} m_{44} + \left< y p_y \right>_f m_{31} m_{44} \\ & - \left< y y \right>_f m_{41} m_{44}  =0
 \\\\
& -\left< p_x p_x \right>_f m_{13} m_{14} + \left< x p_x \right>_f m_{14} m_{23} + \left< x p_x \right>_f m_{13} m_{24} - \left< x x \right>_f m_{23} m_{24} - 
 \left< p_x p_y \right>_f m_{14} m_{33} \\ & + \left< x p_y \right>_f m_{24} m_{33} - \left< p_x p_y \right>_f m_{13} m_{34} + \left< x p_y \right>_f m_{23} m_{34} - 
 \left< p_y p_y \right>_f m_{33} m_{34} + \left< y p_x \right>_f m_{14} m_{43}  \\ & - \left< x y \right>_f m_{24} m_{43} + \left< y p_y \right>_f m_{34} m_{43} + 
 \left< y p_x \right>_f m_{13} m_{44} - \left< x y \right>_f m_{23} m_{44} + \left< y p_y \right>_f m_{33} m_{44}\\ &  - \left< y y \right>_f m_{43} m_{44} = 0
\end{align*}
 The beam covariances, $\left< ...\right>_f$, and the transfer matrix elements $\{m_{11},m_{21},m_{31},m_{41},m_{13},$ $m_{23},m_{33},m_{43}\}$ are measured by aperture scans. The unknown variables are then the transfer matrix elements $\vec{X} =\{m_{12},m_{22},m_{32},m_{42},m_{14},m_{24},m_{34},m_{44}\}$. This system of equations can be represented by $\vec{f}(\vec{X}) = 0$, where each element of $\vec{f}(\vec{X})$ is a function representing the left-hand side of an equation. 

Given that there are 9 equations and 8 unknown variables, this is an overdetermined system of equations. As such, we can iteratively solve for the unknown transfer matrix elements $\vec{X}$ with the Gauss-Newton steps,
\begin{align*}
    \vec{X}^{(i+1)} = \vec{X}^{(i)} - \mathbf{J}(\vec{X}^{(i)})^{\dag } \vec{f}(\vec{X}^{(i)}),
\end{align*}
where $\mathbf{J}(\vec{X}^{(i-1)})^{\dag }$ is the Moore-Penrose pseudo-inverse of the Jacobian matrix of $\vec{R}$ evaluated at the current value of $\vec{X}$. Typically, the Gauss-Newton steps converge to a solution rapidly within a few steps, with an initial guess of $\vec{X}^{(0)} = 0$. 
\newpage
\section{Cathode Fabrication and Growth}
\begin{figure}[h!]
    \centering
    \includegraphics[width=1\linewidth]{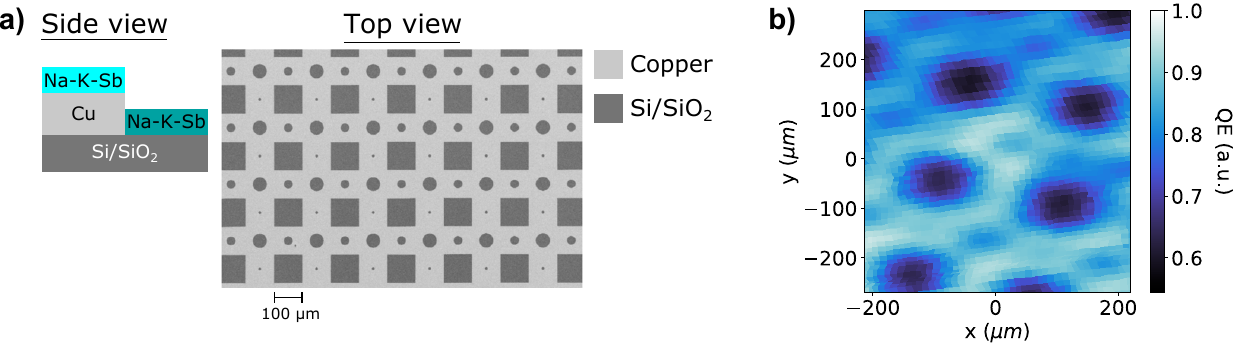}
    \caption{\textbf{a)} Top view shows an SEM image of the substrate. Light grey regions represent copper, and dark grey regions indicate Si/SiO$_2$. The side view shows the layers of the cathode with a thin film of Na-K-Sb grown on top of the substrate. \textbf{b)} QE map of a part of the cathode. Areas of low QE correspond with the regions of the substrate that consist of only Si/SiO$_2$.}
    \label{fig:cathode}
\end{figure}

Our patterned cathodes were made using photolithography at the Cornell NanoScale Facility (CNF), using an image-reversal, lift off procedure. An n-doped silicon (Si) wafer has thin positive resist layer added, which is then patterned in a GCA AutoStep 200 DSW i-line Wafer Stepper. This tool uses 365 nm light to achieve features as small as 400–600 nm. This pattern is then image reversed by exposing it to hot ammonia gas, causing the previously positive resist to behave like a negative resist during development, leaving undercuts appropriate for later liftoff. A layer of copper is then evaporated onto the wafer, and the remaining resist is lifted off, leaving a patterned later of copper on silicon. The resulting patterned substrate is shown in Fig.\,\ref{fig:cathode}a. 


This patterned wafer is then used as the substrate for Na-K-Sb cathode growth in a molecular beam epitaxy (MBE) growth system. In the MBE system, the substrate is annealed to 500°C for 15 minutes to remove any organics from the surface. The growth process employs codeposition of Na, K, and Sb using effusion cells. The rate of fluxes is chosen to target the stoichiometry Na$_2$KSb. During deposition, the QE is monitored using a 532 nm laser and measuring the drain current from the electrically floated sample holder, biased at 73 V. The codeposition was ended when the QE saturated, at 30 minutes. The estimated thickness of the deposition is 14.6 nm, based on measured source fluxes and assumed stoichiometry. Following growth, a spectral response is taken to determine the QE at different wavelengths. These measurements are insensitive to the substrate's micro-scale patterns due to the laser spot size. The QE was measured to be 9.29\% at 400 nm and 1.88\% at 530 nm. 

After the Na-K-Sb deposition, we scanned a 405 nm laser, with a spot size between 20 to 30 $\mu$m rms, across the cathode while measuring the beam current, which yields a spatial map of the cathode QE in Fig.\,\ref{fig:cathode}b. The QE map clearly shows that the QE is greater for the Na-K-Sb grown on copper compared to Si/SiO$_2$.

\newpage
\section{Measuring Initial Beam Size}
\begin{figure}[h!]
    \centering
    \includegraphics[width=1\linewidth]{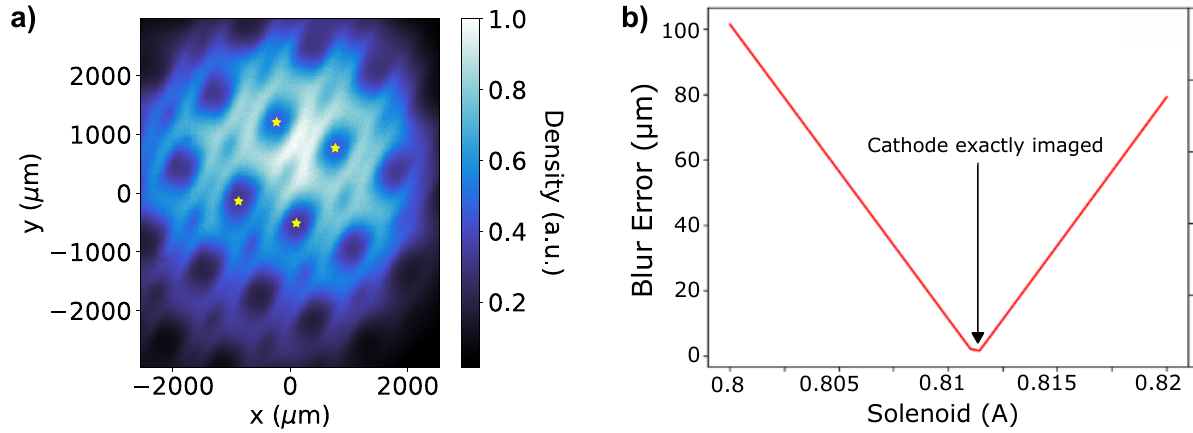}
    \caption{\textbf{a)} Detector image of the beam when the beamline is imaging the cathode spatial distribution of the beam onto the detector. Yellow stars mark the position of measured features in the beam. \textbf{b)} Simulation of the beam size, or blur error, at the detector for a point source at the cathode while varying the solenoid focusing. The cathode is exactly imaged onto the detector when the beam size, or blur error, is zero.}
    \label{fig:MCPimaging}
\end{figure}

We measure the initial beam size by first attempting to image the source spatial distribution of the beam at the cathode onto the detector, as shown in Fig. \ref{fig:MCPimaging}a. This aims to achieve a beamline transfer matrix where $m_{12} = m_{14} = m_{32} = m_{34} = 0$, such that the momentum of the emitted electrons has no effect on their position at the detector. In this exact cathode imaging condition, the measured spatial distribution of the beam is therefore simply a linear transform of its source spatial distribution. This transformation can be easily measured by linearly fitting the position of features in the beam, such as the rectangular dots in Fig. \ref{fig:MCPimaging}a, to the known positions of corresponding features on the cathode, as measured by SEM in Fig. \ref{fig:cathode}a.

However, we are likely not exactly at the cathode imaging condition, so $m_{12} = m_{14} = m_{32} = m_{34} = \epsilon$, where $\epsilon$ is small. Therefore, we need to understand the effect of this error on the measured spatial distribution.
For simplicity, we assume cylindrical symmetry about the longitudinal direction and look at the 2D case, where we only consider $x$ and $p_x$. The rms beam size measured at the detector, $\sigma_{xf}$, is related to the rms beam size, $\sigma_{xi}$, and rms momentum spread, $\sigma_{pxi}$, at the source by,
\begin{align*}
    \sigma_{xf}^2 = m_{11}^2 \sigma_{xi}^2 + m_{12}^2\sigma_{pxi}^2
\end{align*}
Note that we assume zero position-momentum correlation at the source. The latter term can be expressed as $m_{12}^2\sigma_{pxi}^2 = \epsilon^2\sigma_{pxi}^2$ near the cathode imaging condition, and can be understood as an error term that Gaussian blurs the measured spatial beam distribution. 

Using General Particle Tracer codes, we can simulate the blur error while varying the solenoid focusing near the cathode imaging condition, which has zero blur error as $\epsilon \rightarrow 0$, as shown in Fig. \ref{fig:MCPimaging}b. The range of solenoid focusing shown here is chosen to match the range of solenoid focusing where we experimentally see sharp, recognizable features in the beam that match the cathode features. We see that within this range of solenoid focusing, where the cathode might be imaged onto the detector, the largest blur error is roughly $\epsilon\sigma_{pxi} \approx 100$ $\mu$m. Therefore, we can ensure that this is a negligible effect on the measured beam size by using large enough laser spot sizes such that $\sigma_{xf}^2 >> (100 \text{ } \mu \text{m})^2$.

\newpage
\section{Calculating Beam Covariances and Emittance}
\begin{figure*}[h]
    \centering
    \includegraphics[width=1\linewidth]{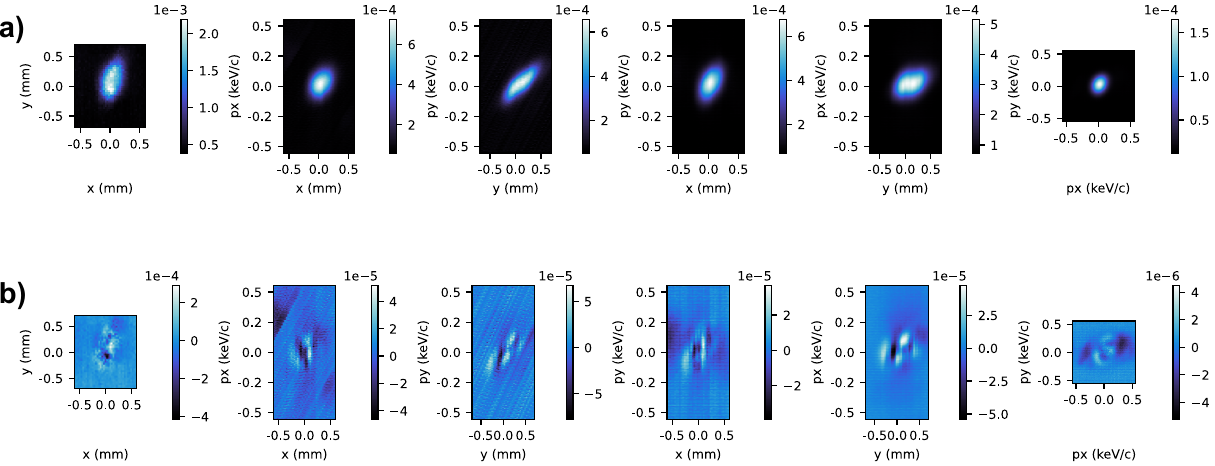}
    \caption{\textbf{a)} Example of 2D projections of a measured 4D phase space. \textbf{b)} Residuals of fitting the 2D projections in \textbf{a)} to a 2D rotated supergaussian.}
    \label{fig:phase-space-fitting}
\end{figure*}
The 4D emittance is calculated from the beam covariance matrix,
\begin{align*}
    \epsilon = \frac{1}{m_e c} \left| \begin{matrix}
        \left< xx\right> & \left<xp_x \right> & \left<xy \right> & \left<xp_y \right> 
        \\
        \left<xp_x \right> & \left<p_xp_x \right> & \left<yp_x \right> & \left<p_xp_y \right> 
        \\
        \left<xy \right> & \left<yp_x \right> & \left<yy \right> & \left<yp_y \right> 
        \\
        \left<xp_y \right> & \left<p_xp_y \right> & \left<yp_y \right> & \left<p_yp_y \right> 
    \end{matrix} \right | ^{\frac{1}{4}}
\end{align*}
To obtain the beam covariances, we fit the 2D projection of a 4D phase space, shown in Fig. \ref{fig:phase-space-fitting}a, to a 2D rotated supergaussian of form,
\begin{align}
    f(u,v) & = A \text{ exp} \left [ -\left(\frac{U^2}{2 \sigma_u^2} + \frac{V^2}{2 \sigma_v^2}  \right)^B \right] + C
    \\
    \begin{pmatrix}
        U \\ V
    \end{pmatrix} & =  \text{Rot}(\theta) \begin{pmatrix}
        u \\ v
    \end{pmatrix} = \begin{pmatrix}
        \text{cos } \theta & -\text{sin } \theta \\ \text{sin } \theta & \text{cos } \theta 
    \end{pmatrix} \begin{pmatrix}
        u \\ v
    \end{pmatrix},
\end{align}
where $u$ and $v$ represent the axes, $\{x, p_x, y, p_y\}$, of the fitted 2D phase space. We can check the residuals of the fit, shown in Fig. \ref{fig:phase-space-fitting}b, to determine whether the fits are of good quality. The beam covariances are then calculated from the fitted 2D supergaussian, where the background term $C$ is set to zero. 

\newpage
\section{Source 4D Phase Space Noise Analysis}
\begin{figure*}[h]
    \centering
    \includegraphics[width=1\linewidth]{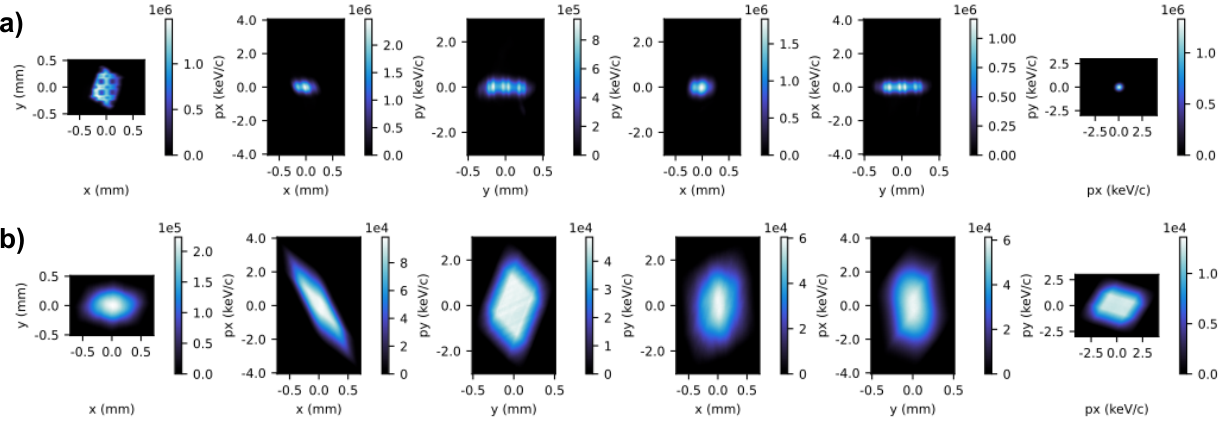}
    \caption{\textbf{a)} 2D projection of the reconstructed source 4D phase space of a beam. \textbf{b)} Background noise of the 2D phase spaces in \textbf{a)} due to camera background noise.}
    \label{fig:source-background-noise}
\end{figure*}
The background noise of the reconstructed source 4D phase space, accounting only for camera background noise, is fully shown in Fig \ref{fig:source-background-noise}b. Comparing it to the source 4D phase space itself (Fig. \ref{fig:source-background-noise}a), the magnitude of the background noise is relatively small, where the signal-to-noise ratio is within a few percent, except for the $x$-$y$ position phase space, where the background noise magnitude is roughly $10\%$ of the signal.